\documentclass[12pt,amsmath,amssymb,superscriptaddress,longbibliography]{article}
\usepackage[utf8]{inputenc}
\usepackage[backend=biber,style=nature]{biblatex}
\usepackage[a4paper, total={6.5in, 8.5in}]{geometry}
\usepackage{authblk}
\usepackage{graphicx}
\usepackage{physics}
\usepackage{siunitx}
\usepackage[pagewise]{lineno}
\usepackage{xcolor}
    
\title{Boundary-Shape Driven Transitions in Vortex and Oscillatory Dynamics of Confined Epithelial Cells}
\author[1,3]{Ryo Ienaga}\affil[1]{Department of Chemical Engineering, Kyoto University, Nishikyo-ku, Kyoto 615-8510, Japan}
\author[2,3]{Kazuyuki Shigeta}\affil[2]{Department of Physics, Kyushu University, Nishi-ku, Fukuoka 815-0395, Japan}
\author[1,3]{Tatsuya Fukuyama}
\author[1,*]{Kazusa Beppu}
\author[1,*]{Yusuke T. Maeda}
\affil[3]{These authors equally contributed to this work}
\affil[*]{corresponding address: kazusa.beppu@cheme.kyoto-u.ac.jp; maeda@cheme.kyoto-u.ac.jp}

\begin{document}

\maketitle

\begin{abstract}Controlling the collective motion of epithelial cell populations is fundamental for understanding multicellular self-organization and for advancing tissue engineering. Under spatial confinement, cells are known to exhibit either vortex rotation or oscillatory motion depending on boundary geometry, but the mechanisms governing transitions between these states remain unclear. Here, we investigated the collective motion of MDCK cells confined within a doublet circular boundary, where the confinement aspect ratio, defined as the distance between the centers of two circles relative to their radius, can be tuned by varying the degree of overlap. When the overlap is large, cells form a stable vortex. Increasing the confinement aspect ratio destabilizes this vortex and induces oscillatory motion characterized by periodic reversals of migration direction, before ultimately transitioning into disordered dynamics. To elucidate the underlying mechanism, we developed simulations of self-propelled particles incorporating local alignment (LA) and contact inhibition of locomotion (CIL). The model successfully reproduced the experimentally observed transitions from vortices to oscillatory motion and further revealed that an appropriate balance between LA and CIL is critical for stabilizing vortex pairs with velocity reversals. Our findings demonstrate that the confinement aspect ratio serves as a minimal control parameter governing transitions in the collective dynamics of epithelial monolayers.
\end{abstract}

\maketitle

\section{Introduction}
Epithelial cell populations frequently display disordered but locally polarized and ordered motion on two-dimensional substrate \textit{in vitro} and wound healing process \textit{in vivo} \cite{blanch2018turbulent, lin2021energetics,brugues2014forces,basan2013alignment, aoki2017propagating}. Such collective dynamics are often characterized as \textit{active turbulence}, in which the motion resembles turbulence yet retains a hidden ordered structure \cite{alert2020physical, alert2022active,ramaswamy2010mechanics, marchetti2013hydrodynamics}. Active turbulence is a universal phenomenon not only in epithelial monolayers but also in dense bacterial suspensions \cite{wensink2012meso,dunkel2013fluid}, in active fluids composed of molecular motors and microtubule cytoskeletons \cite{needleman2017active,sanchez2012spontaneous,tan2019topological}. Importantly, this turbulent-like motion exhibits velocity correlations with a characteristic spatial scale, reflecting emergent order such as vortex-like structures. Despite differences in interaction details (polar interactions or nematic interactions), elucidating the mechanisms by which uniformly aligned states become unstable and give rise to turbulence remains a fundamental challenge, with broad implications ranging from the physics of active matter to biophysics.

To uncover the hidden ordered structures within turbulent dynamics, it is necessary to identify ways to stabilize the ordered motion. Owing to the characteristic length scale inherent in active turbulence, collective motion can be effectively controlled by geometric boundaries. For example, narrow channel boundaries drive cells to migrate in a single direction, and as cells move along walls or ridges, the boundary shape dictates the direction of migration \cite{vedula2012emerging,deforet2014emergence,duclos2017topological,duclos2018spontaneous,pricoupenko2024src}. 

A representative approach for imposing geometric boundaries is micropatterning technology, which permits cell adhesion only within predefined regions while preventing adhesion elsewhere. This technique enables systematic analysis of collective motion patterns under designed geometric boundaries. It is well established that vortex motion can be stabilized when the confinement geometry is circular with a size comparable to the velocity correlation length scale \cite{deforet2014emergence,segerer2015emergence,yu2021spatiotemporal}. Furthermore, previous studies have shown that not only two-dimensional planar surfaces but also three-dimensional cylindrical and wrinkled surfaces induce orientation patterns that correspond to the underlying geometric structures \cite{yevick2015architecture,rupprecht2017geometric,xi2017emergent,shigeta2022collective}. In active fluids, including epithelial cell monolayers with polar interactions and active nematic cell populations \cite{guillamat2022integer,harmand20213d,latorre2018active,ienaga2023geometric}, as well as bacterial populations \cite{wioland2013confinement,wioland2016ferromagnetic,beppu2017geometry, beppu2021edge,beppu2024geometric,nishiguchi2018engineering} or active cytoskeletons \cite{wu2017transition,opathalage2019self,hardouin2019reconfigurable,araki2021controlling,hardouin2022active}, the use of geometrically defined boundaries has become an established method for controlling the collective ordering of active matter.

Geometric control provides a minimal, tunable approach to regulate active turbulence since the ordered collective motions that emerge under geometric confinement are not limited to vortices. Vortex motion is stabilized under circular boundaries, where symmetry is preserved and the confinement is isotropic. In contrast, when the confinement becomes anisotropic, different modes of motion are expected to be stabilized. Indeed, when the boundary is rectangular rather than circular, oscillatory motion arises in which the direction of cell migration periodically alternates along the long axis of the confinement \cite{petrolli2019confinement,peyret2019sustained,parmar2025prxl}. Epithelial cells, which maintain strong intercellular adhesion through transmembrane adhesion proteins, migrate collectively while undergoing viscoelastic deformation. The distinct modes of motion observed under anisotropic confinement, such as in rectangular geometries, suggest that the dynamics of cell populations are strongly influenced by boundary anisotropy.  However, how a vortex stabilized under isotropic circular confinement transitions into oscillatory motion under anisotropic boundaries remains unresolved. In particular, it is unknown whether a single geometric parameter can govern pattern transitions in collective dynamics, highlighting the need for new experimental approaches.

Previous work has established vortex stabilization in circular confinement and oscillations in rectangular confinement, but no study has systematically addressed transitions driven by a single geometric parameter. In this study, to address this issue, we designed a confinement geometry consisting of two overlapping circular boundaries. This configuration enables us to investigate the interaction of two vortices within a confined space. The doublet geometry allows the aspect ratio, and thus the degree of anisotropy, to be varied while still preserving the boundary conditions that support vortex formation. We found that cells under circular spatial constraints showed ordered vortex rotation was observed as reported previously. However, as the distance between the centers of the two circles increased, which corresponds to greater anisotropy, the vortex motion gradually transitioned into oscillatory motion along the longitudinal axis, and eventually developed into disordered dynamics. Our findings suggest that the aspect ratio of the confined space is a key parameter governing the modes of collective motion, thereby advancing our understanding of confined epithelial populations and providing insight into the geometric control of active turbulent dynamics.

\section{Materials and Methods}
\subsection*{Cell culture}
The epithelial cells used in this study were MDCK cells (NBL-2, JCRB9029) grown at \SI{37}{\celsius} in a CO${_2}$ incubator with 5.0\% CO${_2}$ and 90\% humidity as previously reported \cite{shigeta2022collective}. The growth medium used was Dulbecco's modified Eagle's medium (11965092, Thermo Fisher Scientific) supplemented with 10\% fetal bovine serum (173012, NICHIREI Biosciences). Cell density was determined using a cell counter (TC20, Bio-Rad). We used a collagen-coated culture dish (35 mm, IWAKI) for the microscopic observation.

\subsection*{Microfabrication and sample preparation}
We performed microfabrication of the cell adhesion area using reverse microcontact printing of PDMS (Sylgard 184, Corning) as used in the previous study\cite{ienaga2023geometric}. First, we fabricated a PDMS block with the surface pattern of the adhesion region and then treated it with plasma surface treatment to adhere it to a culture dish. The ethanol solution containing 0.1\% MPC polymer (Lipidure, NOF Corporation) was poured by capillary force into the narrow space between the PDMS block and the culture plate to form a nonadhesive area. The PDMS block was then peeled off to open a cell adhesion area. The photomask for designing the PDMS blocks was purchased from MITANI Micronics.

\subsection*{Microscopy and data processing}
Time-lapse recordings of the collective motion of the MDCK cell monolayer were obtained at 5-minute intervals for 72 hours. An inverted fluorescence microscope (IX73, Evident) equipped with a CMOS camera (Zyla, Andor Technology) was used with a phase-contrast imaging setup. The temperature of the microscope stage was maintained at \SI{37}{\celsius}, and the CO$_{2}$ concentration was maintained at 5.0\% using a stage-top cultivation chamber (STXG-IX3WX, TOKAI-Hit). The microscope stage was motorized (BIOS-Light, Optosigma), and multipoint acquisition was performed using Metamorph software. All microscopic images were analyzed, and image processing and calculation of the autocorrelation function were performed using MATLAB software. In particular, we carried out particle image velocimetry (PIV) analysis using ImageJ software with the iterative PIV plugin to obtain the velocity field of the collective motion of MDCK cells. 

\subsection*{Numerical simulation}
To investigate geometry-dependent collective migration, we developed a self-propelled particle model confined within a doublet circular boundary \cite{grossman2008emergence,lin2018dynamic}. Each particle $i$ is described by its center-of-mass position $\mathbf{r}_i(t)$ and polarity vector $\mathbf{q}_i(t)=(\cos\theta_i,\sin\theta_i)$, where $\theta_i(t)$ denotes the polarity angle. The overdamped equation of motion is given by
\begin{equation}
\frac{d\mathbf{r}_i}{dt}
= v_0 \mathbf{q}_i(t)
+ \sum_{j \in A_i} \mathbf{F}_{ij}^{\text{rep}}
+ \sum_{\text{wall}} \mathbf{F}_{i\text{w}},
\end{equation}
where $v_0$ is the self-propulsion speed, and $A_i$ denotes the set of neighboring particles identified by a Voronoi tessellation. Interactions are evaluated only for neighbors within a cutoff distance of $3\sigma_r$.

Intercellular interactions consist solely of short-range soft repulsion. The relative displacement between particles $i$ and $j$ is defined as $\mathbf{r}_{ij}=\mathbf{r}_i-\mathbf{r}_j$, and the repulsive force is modeled as
\begin{equation}
\mathbf{F}_{ij}^{\text{rep}} = - \nabla_{\mathbf{r}} U_r(\mathbf{r}_{ij}), \quad
U_r(\mathbf{r}_{ij}) = U_0 \exp\!\left(-\frac{|\mathbf{r}_{ij}|^2}{2\sigma_r^2}\right),
\end{equation}
where $\sigma_r$ sets the characteristic range of repulsion.  

The confining boundary condition is implemented by a steric repulsion acting when the particle-boundary distance is smaller than $1.5\sigma_r$, modeled as $\mathbf{F}_{i\text{w}} = -\nabla_{\mathbf{r}} U_r(\mathbf{r}_{i\text{w}})$ with $\mathbf{r}_{i\text{w}}=\mathbf{r}_i-\mathbf{r}_\text{w}$ and $\mathbf{r}_\text{w}$ denoting the nearest boundary point. In addition, to prevent particles from crossing the boundary under strong compression, a harmonic potential directed toward the circle center is imposed at the boundary edge.

The polarity of each particle evolves under the competing effects of local alignment (LA) and contact inhibition of locomotion (CIL) \cite{lin2018dynamic}. The polarity angle $\theta_i(t)$ changes according to
\begin{equation}
\frac{d\theta_i}{dt}
= \frac{\mu_a}{n_i}\sum_{j \in A_i}\sin(\theta_j^{\text{vel}}-\theta_i)
+ \frac{\mu_c}{n_i}\sum_{j \in A_i}\sin(\theta_j^{\text{rel}}-\theta_i),
\end{equation}
where $\mu_a$ and $\mu_c$ are the coupling strengths of LA and CIL, respectively, $\theta_j^{\text{vel}}$ denotes the velocity angle of particle $j$, $\theta_j^{\text{rel}}$ is the angle of the relative position vector $\mathbf{r}_{ij}$, and $n_i$ is the number of neighboring cells of particle $i$. For simplicity, stochastic noise is neglected.

The system of equations is nondimensionalized by choosing the repulsive interaction length $\sigma_r$ as the unit of length and the relaxation time $\tau=\sigma_r^2/U_0$ as the unit of time, where $U_0$ is the characteristic strength of repulsion. \textcolor{black}{Based on the typical scale of MDCK cells, which have a size of $\SI{20}{\micro\meter}$ and migrate at a velocity of approximately $\SI{0.3}{\micro\meter\per\minute}$, we reasonably set $\sigma_r=\SI{10}{\micro\meter}$ and $U_0=\SI{60}{\micro\meter^2\per\minute}$, which yield $\tau=\sigma_r^2/U_0=\SI{1.67}{\minute}$.} All the simulations are performed with nondimensional parameters: $\hat{v}_0=0.05$, $\hat{\sigma}_r=1$, and $\hat{\mu}_a=0.05$. The radius of the doublet circular boundary is set to $R = 10$. The repulsive strength and area fraction are fixed at $\hat{U}_0=1$ and $\hat{\Phi}=0.6$, respectively, where $\hat{\Phi}=N\pi\sigma_r^2/S$ with $N$ the number of particles and $S$ the boundary area, unless otherwise specified. In this study, the CIL strength $\mu_c$ is varied in the range $0.1 \leq \mu_c \leq 1$ to elucidate its influence on vortex formation and oscillatory dynamics.

Simulations were initialized with random particle positions and orientations and integrated for $10^5$ time steps using the Euler method ($\Delta t=0.1$), with data sampled every 100 steps.

\section{Results}

\begin{figure*}[tb]
\begin{center}
\includegraphics[width=17cm]{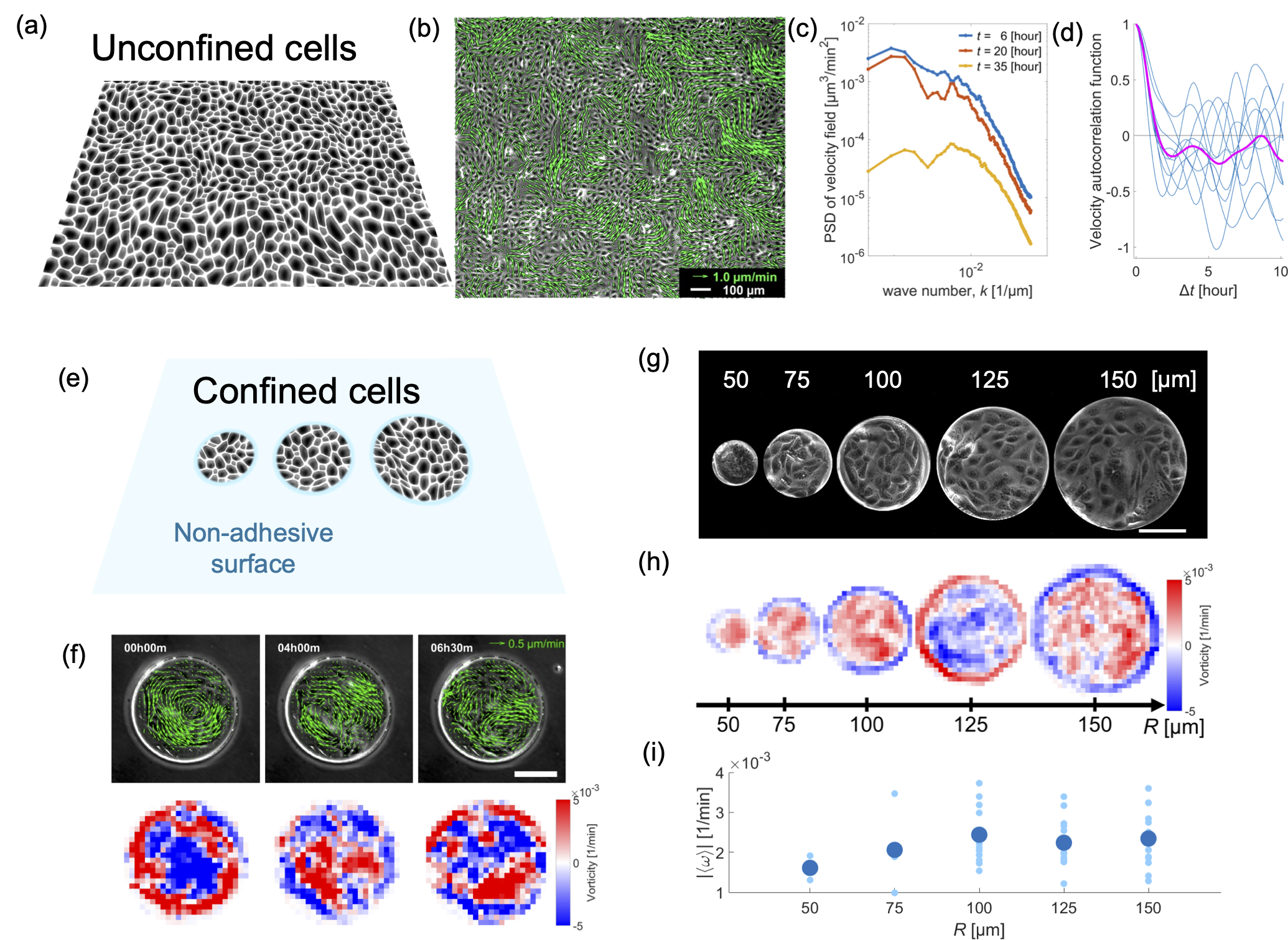}
\caption{{\footnotesize\textbf{Collective motion of MDCK cell populations under flat two-dimensional plane and circular confinement conditions.} 
(a) Schematic illustration of unconfined cells. 
(b) Collective motion of the MDCK epithelial cell population in the unconfined state. 
(c) Power spectral density (PSD) of the velocity field of MDCK cells. 
(d) Temporal correlation of collective motion in unconfined MDCK cells. The solid purple line represents the averaged correlation function.
(e) Schematic illustration of confined cells under a circular spatial constraint. 
(f) Top: phase-contrast image of circularly confined MDCK cells superimposed on the velocity field. Bottom: corresponding vorticity field of the collective motion. Scale bar: \SI{100}{\micro\meter}. 
(g) Phase-contrast images of MDCK cells confined within circles of various radii $R$. 
(h) Size dependence of the vorticity field of the collective motion. The color map stands for the vorticity field of the confined MDCK cells. 
(i) Absolute vorticity of confined MDCK cells, $|\langle \omega \rangle|$, as a function of the confinement radius $R$.}}
\label{fig1}
\end{center}
\end{figure*}

We used MDCK cells to study collective motion, showing the transitions of ordered patterns under geometric constraints. First, as a control, we measured collective motion on a two-dimensional surface of a culture dish without any spatial constraints. The collective motion of MDCK cells was measured by Particle Image Velocimetry (PIV) as the cells grew and reached 100\% confluence in the culture dish (Fig. \ref{fig1}(a)). The velocity field $\mathbf{v}(\mathbf{x},t)$ with the position $\mathbf{x}=(x,y)$ varied spatio-temporally (Fig. \ref{fig1}(b) and Movie S1). To analyze the characteristic structures in the collective motion, we calculated the power spectrum of the velocity field at various time points ($t$=6, 20, 39 hours) as used in previous studies \cite{blanch2018turbulent,deforet2014emergence,petrolli2019confinement,shigeta2022collective}. We found that the power spectrum density (PSD) of the velocity field shows a peak at $l_v$=\SI{300}{\micro\meter} at $t=6$ hours and 20 hours (Fig. \ref{fig1}(c)). This peak of PSD means the characteristic length scale reflecting the vortex motion behind the turbulent dynamics \cite{deforet2014emergence,yu2021spatiotemporal}. At a later time ($t=39$ hours), the PSD became lower in magnitude because the collective motion was jammed (a frozen state in motion) as the cell density increased. In the following analysis, we examined the turbulent state with high motility. We also analyzed the temporal autocorrelation function of $\mathbf{v}(\mathbf{x},t)$ by using $C(\Delta t)=\frac{\langle \mathbf{v}(r,t + \Delta t)\cdot\mathbf{v}(r,t) \rangle_{r,t}}{\langle \mathbf{v}^2 \rangle_{r,t}}$ with radial distance $r$ and the ensemble average $\langle\cdot\rangle_t$. $C(\Delta t)$ at various positions shows periodic peaks around $\Delta t \approx $4 hours, reflecting a hidden ordered dynamics even in the turbulent-like collective motion. (Fig. \ref{fig1}(d)). The correlation length of the velocity field of this collective motion shows that the MDCK cells exhibit collective motion in a uniform direction and serves as a typical length standard for controlling motion by specifying the boundary shape.

The typical structure of the velocity field extracted by power spectrum density is a vortex pattern, and numerous reports in the past have shown that this gives a circular boundary shape \cite{deforet2014emergence}. However, since the length of the velocity correlation varies depending on the reported case, we investigated whether collective motion of vortices appears by giving a circular boundary based on the data obtained in Fig. \ref{fig1}(c). We examined the collective motion in circular confinement with a diameter of $R =$\SI{150}{\micro\meter}, which is comparable to the velocity correlation length ($l_v/2\approx$\SI{150}{\micro\meter})(Fig. \ref{fig1}(e)). We found that the confined MDCK cells collectively rotate in one direction along the circular boundary, that is, vortex-like motion. This rotational motion caused a velocity reversal, whereby the direction of rotation reverses over time (Fig. \ref{fig1}(f) $t=0$: counter-clockwise rotation, $t=4$ hours: clockwise rotation, Movie S2). This reversal is consistent with what has been observed in past studies \cite{segerer2015emergence,yu2021spatiotemporal}, and the velocity reversal has both left and right rotations at approximately the same probability, indicating that the chirality of the direction of rotation was not significant.

To characterize the vorticity of rotational vortex motion, we defined the mean absolute vorticity within the circular boundary as $|\langle\omega\rangle|$. We varied the confinement radius $R$ from \SI{50}{\micro\meter} to \SI{150}{\micro\meter} (Figs.~\ref{fig1}(g), (h)) and compared the corresponding values of $|\langle\omega\rangle|$ (Fig.~\ref{fig1}(i)). As $R$ increased from \SI{50}{\micro\meter} to \SI{100}{\micro\meter}, the mean vorticity $|\langle\omega\rangle|$ increased, but it then plateaued and remained nearly constant up to \SI{150}{\micro\meter} (Fig.~\ref{fig1}(i)). When the confinement is too small, vortices can form, but rotational motion is restricted, leading to a reduction in vorticity. These results suggest that a confinement size on the order of the velocity correlation length is optimal for capturing the formation of vortex structures at the circular boundary, with $l_v/2\approx$\SI{150}{\micro\meter}.

\begin{figure*}[tb]
\begin{center}
\includegraphics[width=17cm]{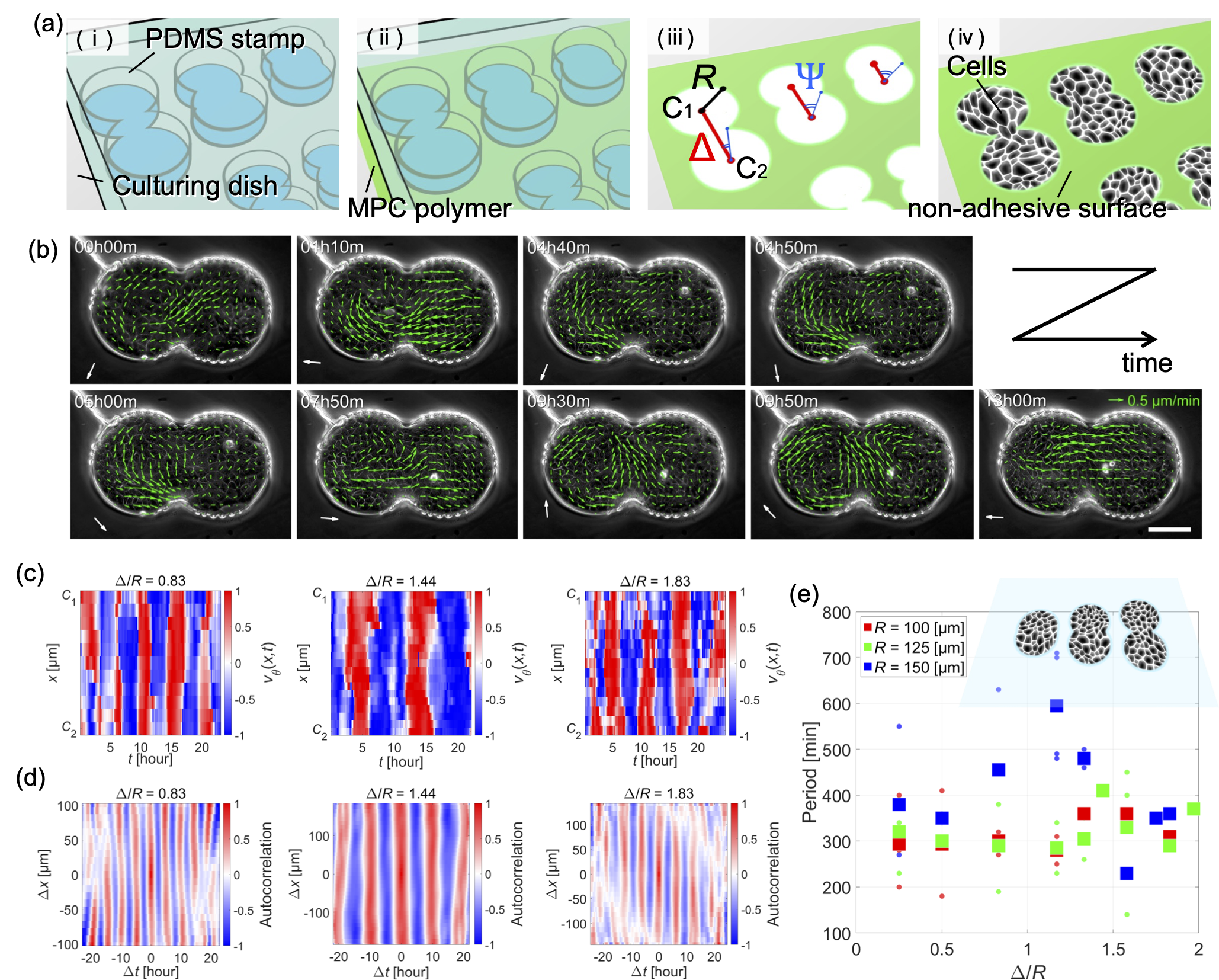}
\caption{{\footnotesize\textbf{Oscillation and velocity reversal in the collective motion of MDCK cells in the doublet circular boundary.}
(a) Schematic illustration of the confinement geometry in the doublet circular boundary. 
(b) Oscillatory collective motion of MDCK cells confined in the doublet circular boundary with $R=\SI{150}{\micro\meter}$ and $\Delta/R=1.44$. Scale bar: \SI{100}{\micro\meter}. The white arrow indicated in the lower left corner stands for the averaged direction of the velocity field.
(c) Periodic velocity reversal of MDCK cells confined within the doublet circular boundary of $R=\SI{150}{\micro\meter}$. The velocity component along the major axis, measured from the center of circle 1 ($C_1$) to the center of circle 2 ($C_2$) as shown in Fig.~\ref{fig3}(a), is plotted over time. Left: $\Delta/R=0.83$; center: $\Delta/R=1.44$; right: $\Delta/R=1.83$. 
(d) Autocorrelation functions of the velocity fields shown in (a). 
(e) Oscillation period of velocity reversal, $k_t^{-1}$, as a function of $\Delta/R$. Data are shown for different confinement radii (red: \SI{100}{\micro\meter}; green: \SI{125}{\micro\meter}; blue: \SI{150}{\micro\meter}). Small dots represent individual experimental data, and filled squares are mean values.}}
\label{fig2}
\end{center}
\end{figure*}

The circular boundary condition can induce vortex rotation accompanied by velocity reversal, yet how interactions between such vortices are modulated by geometric constraints has been little explored. To investigate this, it is useful to design confinement geometries in which the rotational patterns of collective motion change as a result of mutual interactions. With this aim, we introduced a boundary composed of two partially overlapping circles and analyzed the collective motion of epithelial cells within this constrained space (Fig.~\ref{fig2}(a)). The distance between the centers of the two circles is defined as $\Delta$ and the radius of each circle is fixed at $R$=\SI{150}{\micro\meter}. We refer to this geometry as the doublet circular boundary.

In this geometry, changing the center-to-center distance $\Delta$ while keeping the circle size $R$ constant alters the confinement aspect ratio of the spatial constraint for the epithelial cell population. The confinement aspect ratio is defined as the distance between the centers of two circles $\Delta$ relative to their radius $R$, which can be tuned by varying the degree of overlap of two circles. This aspect ratio is expressed as the ratio of two length scales $\Delta/R=\cos\Psi$ where $\Psi$ is the elevation angle that also determines the collision angle of a vortex pair. Thus, $\Delta/R$ serves as a key geometric parameter that governs the strength of vortex interactions within the confined region. We prepared multiple geometric confinement patterns by varying $\Delta/R$ from 0 to 1.97 in this doublet circle boundary, and measured the collective motion of MDCK cells by performing multi-point time-lapse measurements.

As an example, we present an image of the velocity field of collective MDCK cell motion at $\Delta/R$=1.44 (Fig.~\ref{fig2}(b) and Movie S3). Under this condition, although the collective exhibits vortex-like rotation, it is evident that instead of two vortices coexisting simultaneously, one vortex rotates (4 hours 40 min in Fig.~\ref{fig2}(b)) while another appears at a different location and time (9 hours 30 min in Fig.~\ref{fig2}(b)). In addition, cells move along the longitudinal axis, displaying a periodic reversal of direction in an oscillatory manner.

When the cell collective motion was measured under the condition of $\Delta/R = 1.44$, where the aspect ratio was increased, we found that the cells also moved along the longitudinal axis and showed a periodic change of direction in an oscillatory manner (4 hours 10 min and 7 hours 50 min in Fig.~\ref{fig2}(b)). To quantify the oscillatory behavior, we analyzed the time evolution of the velocity component in projection onto a line passing through the center of the two circles $C_1$ and $C_2$ (Fig. \ref{fig2}(a)iii). This allows the sign change of the velocity component to report whether the motion along the long axis direction is leftward or rightward (Fig. \ref{fig2}(b)). These results show that boundary asymmetry transforms rotational vortices into oscillatory motion characterized by periodic velocity reversals.

\begin{figure*}[tb]
\begin{center}
\includegraphics[width=16cm]{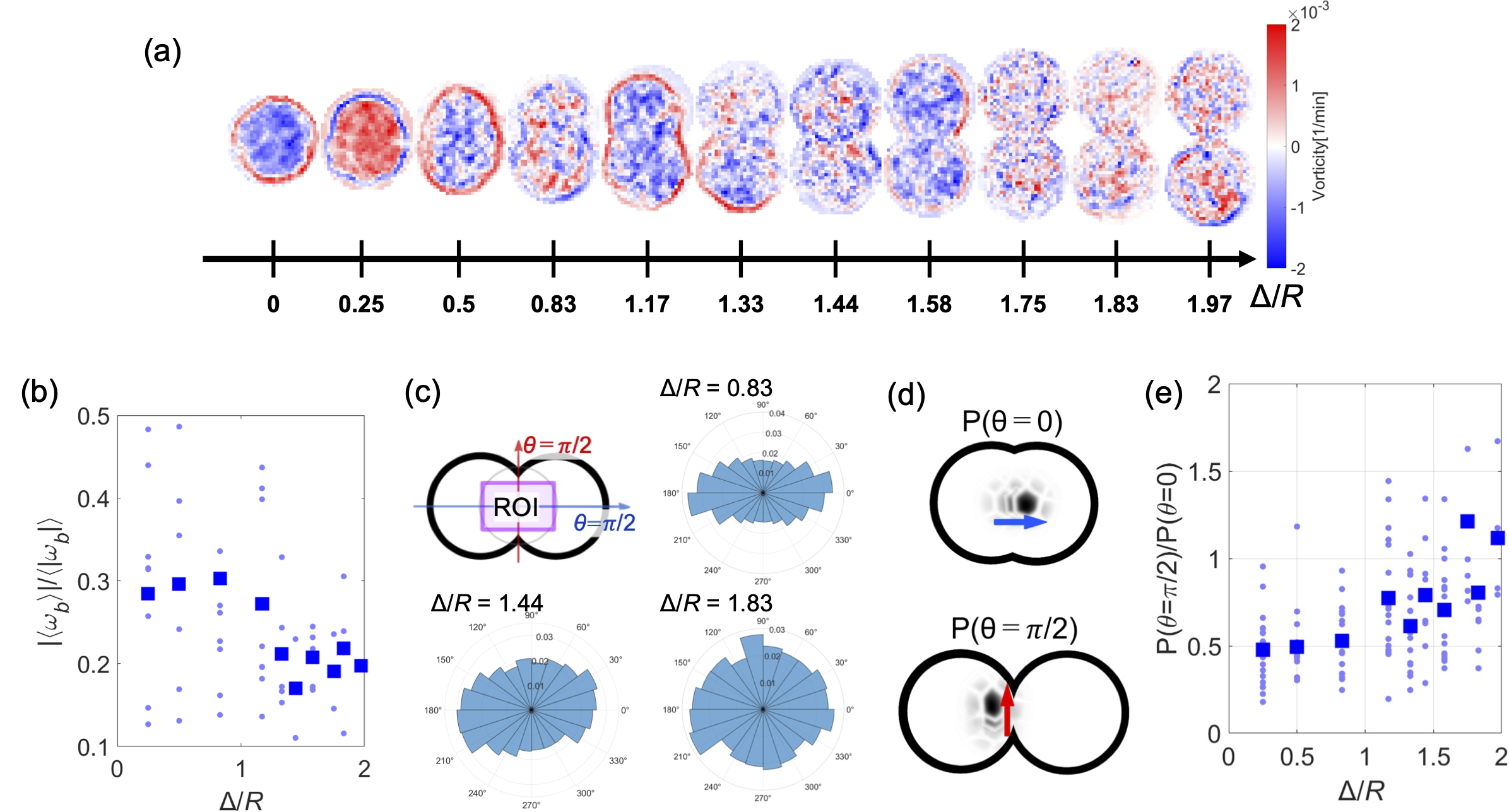}
\caption{{\footnotesize\textbf{Transition from vortex rotation to oscillatory motion of MDCK cells in the doublet circular boundary.} 
(a) Vorticity patterns of MDCK cells confined in the doublet circular boundary with $R=\SI{150}{\micro\meter}$. 
(b) Geometry-dependent change in the vortex order parameter, $|\langle \omega_b \rangle|/\langle|\omega_b|\rangle$, where $\langle \omega_b \rangle$ is the average vorticity within \SI{20}{\micro\meter} of the boundary. The order parameter shows an abrupt change at approximately $\Delta/R \approx 1.33$. Small dots represent individual experimental data, and blue squares are mean values.
(c) Schematic illustration of the analysis of motion direction in the doublet circular boundary. 
The region of interest (ROI) for the analysis is defined (upper left), and velocity orientation $\theta$ is defined as the angle from the geometric center of the doublet circular boundary. Probability distributions of $\theta$ are shown for the confinement aspect ratio of $\Delta/R = 0.83$, $\Delta/R = 1.44$, and $\Delta/R = 1.83$. These distributions are used in the analysis in (e). 
(d) Schematic illustration of cell migration in the long-axis ($P(\theta=0)$) and the short-axis ($P(\theta=\pi/2)$).
(e) Ratio of short-axis to long-axis collective motion.}}
\label{fig3}
\end{center}
\end{figure*}

By calculating the spatio-temporal autocorrelation function of the velocity change, defined as \begin{equation}
    C(\Delta x, \Delta t)=\frac{\langle \mathbf{v}(x + \Delta x,t + \Delta t)\cdot\mathbf{v}(x,t) \rangle_{x,t}}{\langle \mathbf{v}^2 \rangle_{x,t}}.
\end{equation}

We confirmed that clear peaks appeared periodically, with an oscillation period of approximately 6 hours. This periodic velocity reversal was consistently observed across different confinement aspect ratios (Fig.~\ref{fig2}(c) and (d)). To test whether the oscillation period depends on geometry, we analyzed the spatiotemporal autocorrelation function and determined the oscillation period by averaging the peak-to-peak intervals for a range of geometries, $0.25\leq \Delta/R\leq 1.97$. The oscillation period was then plotted against the confinement aspect ratio $\Delta/R$ (Fig.~\ref{fig2}(e)). When $\Delta/R$ was below 1.33, we found that the oscillation period increased as $\Delta/R$ became larger, and this trend was observed regardless of the value of $R$. Around $\Delta/R = 1.33$, the period reached its maximum, after which it decreased again as $\Delta/R$ continued to increase. 

Moreover, to investigate the relationship between vortex dynamics and longitudinal oscillatory motion, we analyzed the vorticity patterns within the doublet circular boundary by varying $\Delta/R$. For small values of $\Delta/R<1.33$, the vorticity exhibited similar values across the region, suggesting that a single vortex motion was present within the doublet circular boundary (Fig.~\ref{fig3}(a)). In contrast, when $\Delta/R \geq 1.33$, the vortex pattern became disordered, and neither a stable vortex structure nor counter-rotating vortex pairs were observed (Fig.~\ref{fig3}(a)). These experimental results show that as $\Delta/R$ increases and the aspect ratio of the confinement space becomes larger, vortex motion transitions from an orderly state to a more irregular vorticity structure.  

To further clarify the confinement aspect ratios under which boundary-aligned vortex motion occurs, we evaluated the vortex order parameter $\frac{|\langle \omega_b\rangle|}{\langle|\omega_b|\rangle}$, where $\langle \omega_b \rangle$ is the average of the vorticity within \SI{20}{\micro\meter} of the boundary. This order parameter approaches 1 when a single vortex or co-rotating vortex pairs are present, and approaches 0 when counter-rotating vortex pairs or random motion dominate. For the confinement aspect ratio $\Delta/R<1.33$, the order parameter was approximately 0.3 (Fig.~\ref{fig3}(b)). However, when the center-to-center distance increased such that $\Delta/R \geq 1.33$, the order parameter decreased to nearly 0 (Fig.~\ref{fig3}(b)). This geometry dependence suggests that co-rotating vortex pairing is favored at smaller $\Delta/R<1.33$.

\begin{figure*}[tb]
\begin{center}
\includegraphics[width=17cm]{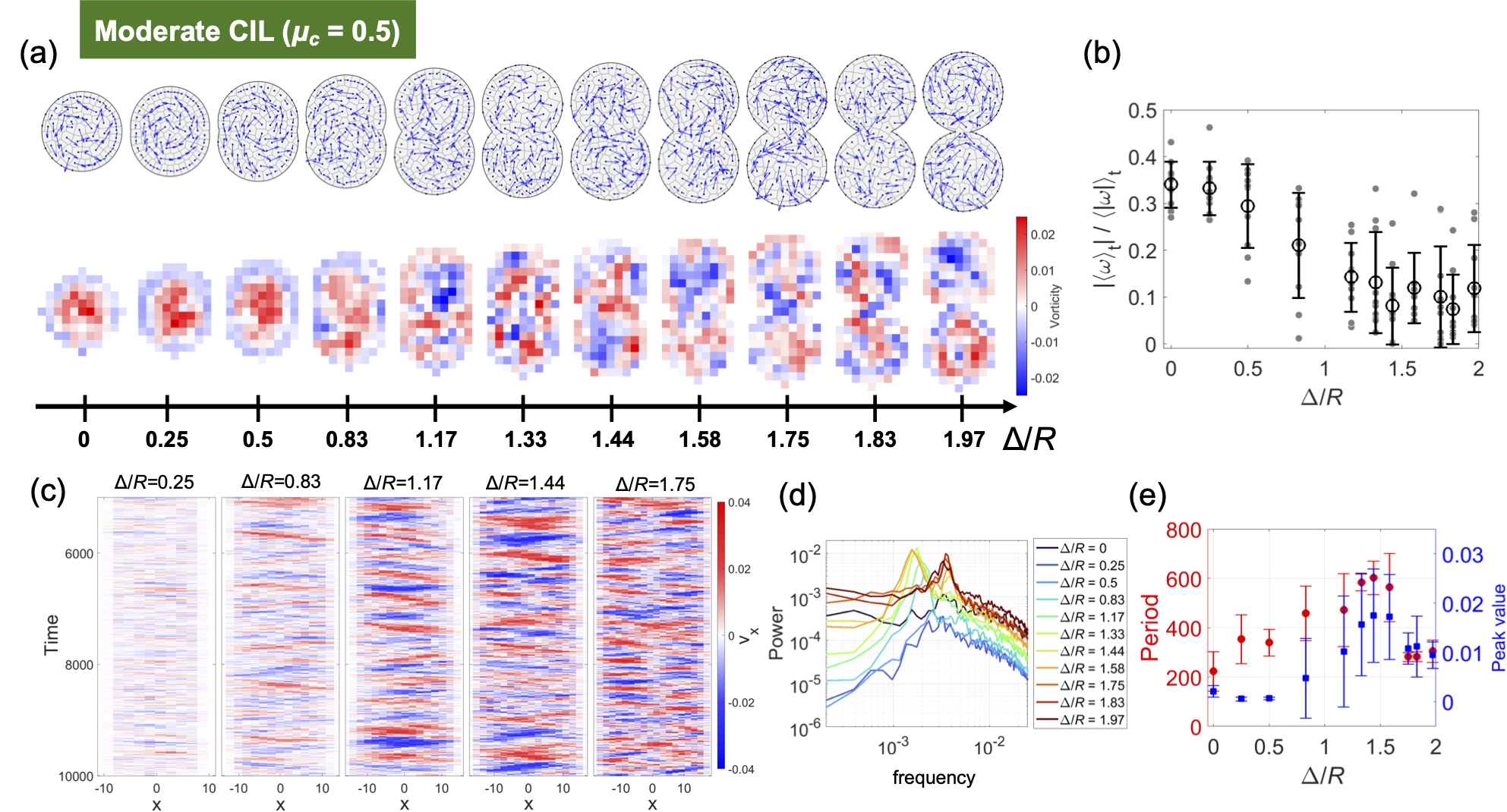}
\caption{{\footnotesize\textbf{Numerical simulation of confined self-propelled particles under LA and moderate CIL interactions ($\mu_c = 0.5$)}. 
(a) Vorticity patterns of self-propelled particles confined within the doublet circular boundary. 
The cell area is divided by Voronoi tessellation. Top: velocity vectors of self-propelled particles; bottom: corresponding vorticity map. 
(b) Vortex order parameter as a function of $\Delta/R$. Small dots represent individual simulation data, each initialized with different random conditions and averaged over the steady-state regime ($T=8\times10^3$ - $10^4$). Open circles and error bars indicate the mean values and standard deviations, respectively. Mean values and standard deviations were obtained from 10 runs with different random initializations, averaged over the steady-state regime ($T = 8\times10^3$--$10^4$). (c) Periodic velocity reversal of confined particles in the doublet circular boundary. The velocity component along the major axis ($v_x$), averaged in the $y$ direction (minor axis) over the ROI between the tips, is plotted as a function of time and $x$ position.
(d) Power spectral density (PSD) of the temporal variation of $v_x$ at various $\Delta/R$ values, averaged over 10 runs with different random initial conditions. 
(e) The oscillation period of velocity reversal, $k_t^{-1}$, and the PSD magnitude of $v_x$ as functions of confinement aspect ratio $\Delta/R$. 
Error bars in (e) represent the standard deviations from 10 runs with different random initializations.}}
\label{fig4}
\end{center}
\end{figure*}

Since co-rotational and counter-rotating vortex pairs are well characterized by the direction of motion in the region between the tips where the two circles overlap. Therefore, we decided to calculate the angular distribution of the velocity, $P(\theta)$, in the central ROI (region of interest) connecting the two peaks (Fig. \ref{fig3}(c)). Under the $\Delta/R = 0.83$ condition, where a single vortex motion appears, the direction of cell motion in the ROI also has sharp peaks at $\theta=0$ and $\theta=\pi$, representing the oscillatory motion of the left and right (Fig. \ref{fig3}(c)). By contrast, for $\Delta/R = 1.83$, the angular distribution of cell motion was unbiased and equally distributed in all directions (Fig. \ref{fig3}(c)). Furthermore, in order to clarify the geometric dependence of the change in the direction of motion, we compared the probability of oscillatory motion $P(\theta=0,\pi)$ in the left-right direction to the probability of vortex pair motion $P(\theta=\pi/2,-\pi/2)$ in the up-down direction (Fig. \ref{fig3}(d)). We defined the ratio of the probability of motion in the minor axis to that in the major axis as $S=\frac{P(\theta=\pi/2,-\pi/2)}{P(\theta=0,\pi)}$. We found that $S$ is close to 0.5 in the confinement aspect ratio of $\Delta/R<1.33$, indicating that the cells prefer motion in the left and right directions (Fig. \ref{fig3}(e)). As $\Delta/R$ increases, around $\Delta/R\approx1.75$, $S$ increases to 1, meaning that cells tend to move in both the long and short axes without geometry-dependent bias. 
 
From these experimental results, we found that epithelial cell populations confined within the doublet circular boundary undergo a transition in their collective motion patterns depending on the confinement aspect ratio $\Delta/R$. As $\Delta/R$ increases, the motion shifts from an ordered vortex rotation to a more irregular pattern. In particular, at larger $\Delta/R$, the fraction of cells moving along the minor axis of the confinement increases and becomes comparable to the fraction moving along the major axis. As a result, the cells tend to adopt random orientations, moving in all directions with no dominant alignment.

To complement the experimental observations of transitions in collective motion, we performed numerical simulations of self-propelled particles confined within the doublet circular boundary under various $\Delta/R$ conditions (see Numerical Simulation in Materials and Methods, Eqs. (1) - (3)). The dynamics of particle polarity, as described in Eq. (3), are governed by local alignment (LA) and contact inhibition of locomotion (CIL). These interactions play distinct roles: LA aligns the polarity of neighboring particles and acts as the primary driver of vortex formation, whereas CIL reorients particle polarity away from neighboring particles, effectively providing a repulsive interaction. The resulting vorticity fields, calculated from the velocity vectors of individual particles, are shown in Fig.~\ref{fig4}(a). For $\Delta/R \leq 1.17$, the particles exhibited collective motion, forming a unidirectional vortex within the confined space. However, as $\Delta/R$ increased further, the vorticity maps gradually became disordered. Rather than splitting into two distinct vortices, particle motion transitioned into increasingly irregular patterns. This transition point was found to occur at approximately $\Delta/R \approx 1.17$.

To quantitatively analyze the geometric dependence observed in the simulations, we evaluated the vortex order parameter $\frac{|\langle \omega_b\rangle|}{\langle|\omega_b|\rangle}$, where $\langle \omega_b \rangle$ is the average vorticity within the spatial constraint. Similarly, in experimental analysis, $\frac{|\langle \omega_b\rangle|}{\langle|\omega_b|\rangle}$ near 1 indicates coherent co-rotating vortices, while values near 0 mean disordered motion. For small $\Delta/R$, the vortex order parameter was on average around 0.3, but it decreased to approximately 0.1 once $\Delta/R > 1.17$ (Fig.~\ref{fig4}(b)). This result, consistent with experimental observations, suggests that as $\Delta/R$ increases, the stability of boundary-aligned vortex rotation diminishes. As a result, even at higher $\Delta/R$, the system does not settle into a stable pair of counter-rotating vortices but instead displays disordered patterns.  

Furthermore, by analyzing the velocity along the line connecting the centers of the two circles (as in Fig.~\ref{fig4}), we detected velocity reversals over a wide range of the confinement aspect ratio, $0.25 \leq \Delta/R \leq 1.87$, indicative of oscillatory motion (Fig.~\ref{fig4}(c) and Movies S4-S8). The reversal of longitudinal velocity $v_x$ along the major axis was consistently observed under these conditions, in agreement with experimental results (Fig.~\ref{fig2}(c) and (d)). To quantitatively analyze the dynamics of oscillatory motion, we calculated the PSD of $v_x$ for $0 \leq \Delta/R \leq 1.97$ (Fig.~\ref{fig4}(d)). For PSD analysis, the Fourier transform was applied to the autocorrelation functions of the velocity time series. We found a peak in the PSD at all values of $\Delta/R$, indicating that the dynamics exhibit a characteristic oscillation period. We then plotted the oscillation period obtained from the PSD as a function of $\Delta/R$ and found that the period increases for $0 \leq \Delta/R \leq 1.44$ (Fig.~\ref{fig4}(e)). At larger $\Delta/R > 1.44$, the oscillation period became shorter (i.e., the frequency of velocity reversals increased). Under these conditions, the amplitude of periodic velocity changes weakened, and the self-propelled particles moved in random directions. This behavior is consistent with the experimental results at $\Delta/R = 1.83$, where the angular distribution of cell velocities was isotropic (Fig.~\ref{fig3}(c) and (d)). \textcolor{black}{It should be noted that the dimensionless period at $\Delta/R \approx 1.4$ is approximately 600, which corresponds to $\SI{1000}{\minute}$ ($600\tau$) and is comparable to the experimental values.}

\begin{figure*}[tb]
\begin{center}
\includegraphics[width=16cm]{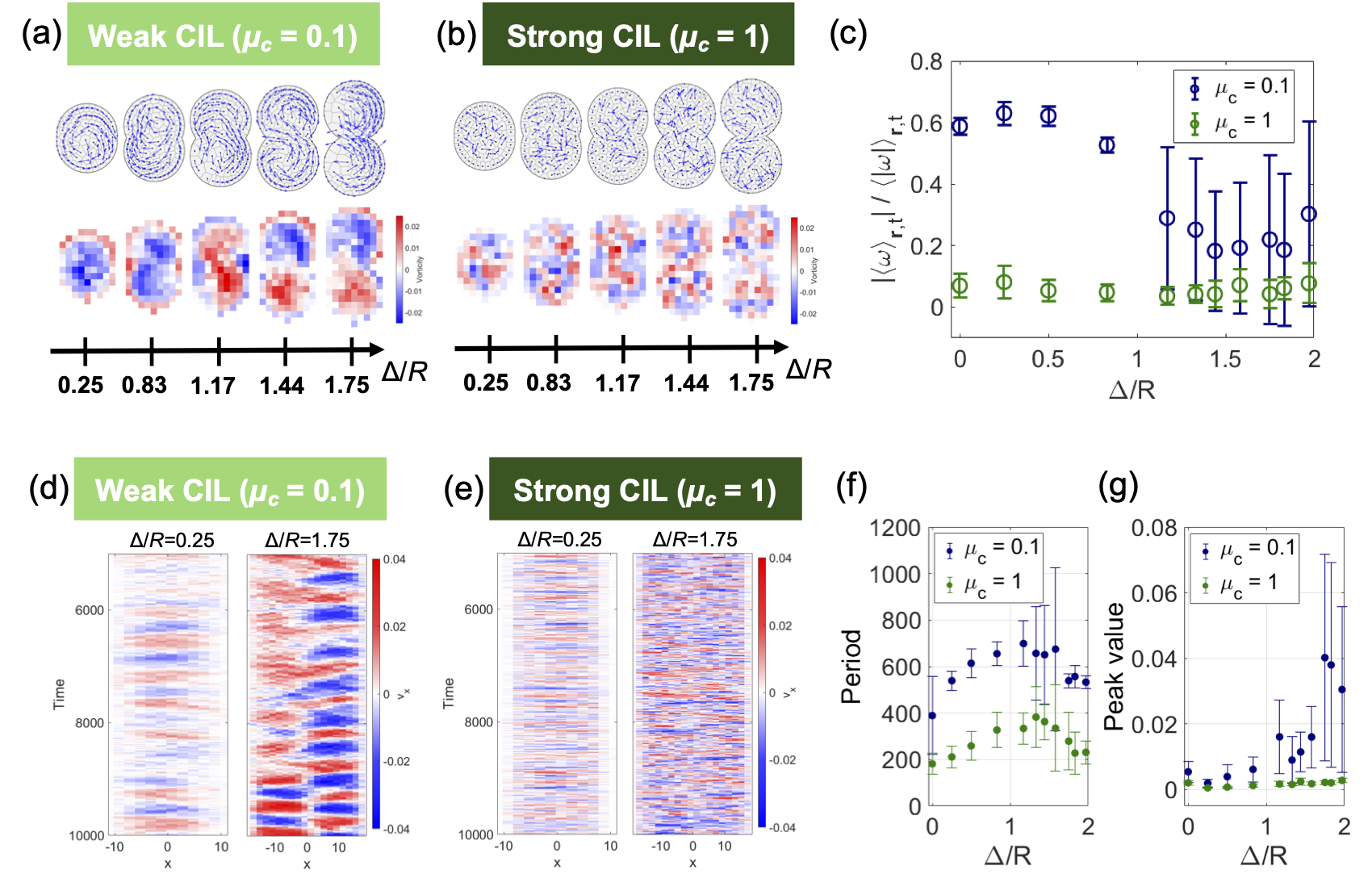}
\caption{{\footnotesize
\textbf{Numerical simulations under weak CIL ($\mu_c = 0.1$) and strong CIL ($\mu_c = 1.0$) conditions}. 
(a) Vorticity patterns of self-propelled particles under weak CIL ($\mu_c = 0.1$) in the doublet circular boundary. Top: velocity vectors of self-propelled particles; bottom: corresponding vorticity map. 
(b) Vorticity patterns of self-propelled particles under strong CIL ($\mu_c = 1.0$) in the doublet circular boundary. 
(c) Vortex order parameter as a function of $\Delta/R$. Blue: weak CIL; green: strong CIL. Mean values and standard deviations were obtained from 10 runs with different random initializations, averaged over the steady-state regime ($T = 8\times10^3$--$10^4$). 
(d) Oscillatory dynamics of confined particles under weak CIL in the doublet circular boundary. The velocity component along the major axis ($v_x$), averaged in the $y$ direction (minor axis) over the ROI between the tips, is plotted as a function of time and $x$ position. 
(e) Irregular dynamics of confined particles under strong CIL in the doublet circular boundary. 
(f) Oscillation period of velocity reversal as a function of the confinement aspect ratio $\Delta/R$. Blue: weak CIL; green: strong CIL. 
(g) Peak PSD of longitudinal velocity oscillations as a function of $\Delta/R$. Blue: weak CIL; green: strong CIL. Error bars in (c), (f), and (g) represent the standard deviations from 10 runs with different random initializations.}}
\label{fig5}
\end{center}
\end{figure*}

By fixing the strength of LA and varying the contribution of CIL, the stability of vortex pairs can be systematically altered. Indeed, in simulations of self-propelled particles, counter-rotating vortex pairs did not appear in geometries with $\Delta/R > 1.33$ when CIL was moderate ($\mu_c = 0.5$). To examine the influence of a weaker CIL, where the effect of LA is more pronounced, we performed additional simulations with $\mu_c = 0.1$ under varying confinement aspect ratios of the doublet circular boundary. Under these conditions, counter-rotating vortex pairs were stably generated when $\Delta/R > 1.33$ (Fig.~\ref{fig5}(a)). We attribute this to the fact that, under weaker contact inhibition, a particle reduces the repulsive force experienced by particles upon contact with neighbors, thereby facilitating directional changes through LA. In contrast, when the contribution of CIL was increased with $\mu_c = 1.0$, vortex motion failed to stabilize even for $\Delta/R < 1.33$, and irregular collective dynamics became dominant (Fig.~\ref{fig5}(b)).  

When the vortex order parameter was calculated under weak CIL conditions, it exhibited only a slight decrease for $\Delta/R > 1.33$, but the onset of vortex order parameter decrease occurred at the same point as in the moderate CIL. Under strong CIL, the vortex order parameter remained close to zero for all confinement aspect ratios, indicating that only disordered collective motion was present. Taken together, these findings highlight that a reduction in CIL allows LA to stabilize vortex pair formation, demonstrating the critical role of alignment interactions in maintaining ordered collective dynamics.

In addition to vortex pair patterns, we also examined the dependence of oscillatory dynamics on the CIL parameter, weak CIL ($\mu_c$=0.1), moderate CIL ($\mu_c$=0.5, Fig. 4), and strong CIL ($\mu_c$=1.0). \textcolor{black}{Under weak CIL conditions, the vortex pair patterns exhibited clear velocity ($v_x$) reversals regardless of $\Delta/R$ (Fig.~\ref{fig5}(d) and Movie S9-S10). This arises because slight heterogeneity among confined cells induces periodic fluctuations of the vortex center along the $y$ direction. Consequently, oscillations of the net $v_x$ appear within the ROI even during vortex rotation, which essentially differs from the oscillatory behavior observed without well-defined vortices under moderate CIL.} In contrast, under strong CIL, periodic velocity fluctuations were not evident, and irregular changes in velocity were observed for both small and large values of $\Delta/R$ (Fig.~\ref{fig5}(e) and Movies S11-S12).  

By analyzing the power spectrum of $v_x$ dynamics, we found that the oscillation period under weak CIL was longer than that under strong CIL (Fig.~\ref{fig5}(f)). Nevertheless, the tendency for the period to reach its maximum around $\Delta/R = 1.44$ was consistent across different CIL strengths. The peak value of the PSD of longitudinal velocity oscillations increased with $\Delta/R$ under weak CIL, whereas under strong CIL it remained significantly lower across all $\Delta/R$ values (Fig.~\ref{fig5}(g)). These results suggest that the contribution of CIL, regulated by $\mu_c$, affects not only the oscillation period but also the range of the confinement aspect ratio over which stable oscillatory patterns can emerge.  

\section{Discussion}

In this study, we investigated the collective motion of epithelial MDCK cells by analyzing their vorticity and velocity fields in confinement geometries with controlled boundary shapes. As reported in previous studies, cells confined within a circular boundary exhibit a vortex-like rotation, and this vortex motion is accompanied by spontaneous directional reversals. When cells are confined within a doublet circular boundary—constructed by superimposing two circles of radius $R$ and varying the center-to-center distance $\Delta$—the collective motion exhibited periodic alternations between leftward and rightward motion along the major axis of the doublet. This directional reversal of the collective flow is reminiscent of previously reported density wave oscillations in MDCK monolayers confined to rectangular geometries \cite{petrolli2019confinement, peyret2019sustained, parmar2025prxl}. In our experiments, the oscillation period was approximately \SI{6}{\hour}, compared to \SI{8}{\hour} in previous reports. Furthermore, it has been shown that when the confinement length along the major axis exceeds \SI{300}{\micro\meter}, the oscillation period becomes independent of system size \cite{petrolli2019confinement}. These findings suggest that the oscillatory dynamics are governed primarily by the geometric aspect ratio, which is the relative lengths of the major and minor axes.

In the experiment, we observed that a single vortex is maintained for $\Delta/R < 1.33$, and periodic directional changes occurred over a broad range of $\Delta/R$. However, when the anisotropy of confinement increased beyond $\Delta/R \geq 1.33$, vortex structures were no longer sustained, and the velocity field became increasingly disordered. Since the oscillatory motion that is characterized by the back-and-forth movement of the entire population exists in broad confinement aspect ratios $\Delta/R$, we infer that the geometric influence controls more prominently the behavior of the cell population along the boundary.  

Our simulations revealed that when orientational interactions of self-propelled particles are governed by both LA and CIL, the emergence of oscillatory dynamics requires an appropriate level of CIL. When CIL is weak, polarity alignment due to LA stabilizes vortex motion, and counter-rotating vortex pairs appear within the doublet circular boundary (Figs.~\ref{fig5}(a) and (d)). In contrast, when CIL is too strong, the collective motion becomes disordered (Figs.~\ref{fig5}(b) and (e)). On the other hand, the oscillatory dynamics have slight dependence on particle density: when the area fraction of particles decreased from $\hat{\Phi} = 0.6$ to 0.4, collective motion became less apparent and both vortices and oscillations disappeared (Fig.~S1), whereas no significant changes were observed at the area fraction of $\hat{\Phi} = 0.8$ (Fig.~S2). Furthermore, increasing the strength of interparticle repulsion did not produce noticeable differences in oscillatory behavior (Fig.~S3). These results indicate that the collective motion of epithelial cells requires LA to stabilize vortex motion, while CIL induces rearrangements that generate oscillatory dynamics.  

Regarding the oscillation period, we observed a linear increase with $\Delta/R$ up to approximately $\Delta/R = 1.44$ in both the experiment ($R = $\SI{150}{\micro\meter}, Fig.~\ref{fig2}(e)) and the simulations (Figs.~\ref{fig4} and~\ref{fig5}). This linear relationship is explained by the fact that a self-propelled particle moving at a constant speed needs time proportional to $\Delta$ to travel from the left end to the right end of the spatial constraints. Thus, the proportionality between the oscillation period and $\Delta$ reflects this simple kinematic relationship. Interestingly, for values of $\Delta/R$ greater than 1.44, the oscillation period decreased sharply. Consistent results from both experiments and simulations near the critical value $\Delta/R = 1.44$ strengthened the link between geometry and dynamic transitions. According to the earlier study \cite{petrolli2019confinement}, when the confinement exceeds about \SI{400}{\micro\meter}, multimodal oscillations appear, but the period becomes constant. The decay of oscillation period for $\Delta/R > 1.44$ is likely a geometric rule of interacting vortex pairs in doublet circle constraints: In active polar fluids of dense bacterial suspensions, it is known that two counter-rotating vortices become stabilized when $\Delta/R$ exceeds $\sqrt{2}$ \cite{beppu2017geometry, beppu2021edge,beppu2024geometric}. In the case of MDCK cells, we speculate that the coexistence of up–down vortex motion with longitudinal oscillation may lead to competing dynamic modes, resulting in irregular motion. \textcolor{black}{These results indicate that the geometry of confinement plays a key role in controlling the robust oscillatory mechanism inherent in epithelial cell populations.}

Moreover, our numerical simulations were developed using a self-propelled particle model. However, the self-propelled particle model neglects cell–substrate traction and mechanical heterogeneity, which may further influence oscillatory dynamics. Whether the observed transitions from vortex to oscillatory and ultimately to disordered states can also be captured by theoretical frameworks such as active polar fluid models \cite{petrolli2019confinement} or active vertex models \cite{fang2022active, barton2017active} remains an open question for future studies. Both the experimental data and the numerical simulations suggest that the oscillation period depends on geometry, with velocity reversals becoming slower near $\Delta/R = 1.44$. Further theoretical refinement, such as incorporating density-dependent polarity or alignment interactions, may be essential to quantitatively explain this geometry dependence of oscillatory motion transformed from the vortex rotation. 

In conclusion, the confinement of interacting vortex pairs via the doublet circle boundary provides a simple yet powerful control parameter for active turbulence. The geometry-dependent transitions observed experimentally, from vortex rotation to oscillatory motion and eventually to disordered dynamics, emerge when the CIL interaction is moderate in such a way that allows reorientation and rearrangement within the confined cell population. This geometric tuning of active turbulence could lead to design principles that bridge epithelial biology and active matter physics.

\section*{Conflicts of interest}

The authors declare no conflicts of interest.

\section*{Acknowledgements}
We thank F. Takabatake and T. Hiraiwa for the fruitful discussion. This work was supported by Grant-in-Aid for Transformative Research Areas (A) (23H04711, 23H04599, 25H04711, 25H04599 to Y.T.M), Grant-in-Aid for Scientific Research (B) (23H01144 to Y.T.M), Grant-in-Aid for Challenging Study (24H01144 to Y.T.M), Nakatani Foundation (to Y.T.M), Nakatomi Foundation (to Y.T.M), JST FOREST Grant JPMJFR2239 (to Y.T.M), Grant-in-Aid for Early-Career Scientists (22K14014 to T.F.), JSPS Core-to-Core Program “Advanced core-to-core network for the physics of self-organizing active matter" (JPJSCCA20230002). R.I. acknowledges the support from JSPS Research Fellowship DC2 (JP25KJ1592) and JST SPRING fellowship.
\printbibliography

\end{document}